\documentclass[twocolumn,english,pre,superscriptaddress,floatfix]{revtex4}
\usepackage{amsmath}
\usepackage[T1]{fontenc}
\usepackage[latin9]{inputenc}
\usepackage{amssymb}
\usepackage{graphicx}
\usepackage{epstopdf}
\usepackage{esint}
\usepackage{times}

\makeatletter

\@ifundefined{textcolor}{}
{%
 \definecolor{BLACK}{gray}{0}
 \definecolor{WHITE}{gray}{1}
 \definecolor{RED}{rgb}{1,0,0}
 \definecolor{GREEN}{rgb}{0,1,0}
 \definecolor{BLUE}{rgb}{0,0,1}
 \definecolor{CYAN}{cmyk}{1,0,0,0}
 \definecolor{MAGENTA}{cmyk}{0,1,0,0}
 \definecolor{YELLOW}{cmyk}{0,0,1,0}
 }

\makeatother

\usepackage{babel}
\begin{document}

\title{A curvature-driven effective attraction in multicomponent membranes}

\author{Matthew F. Demers}
\affiliation{Department of Engineering Sciences and Applied Mathematics, Northwestern
University, Evanston, IL 60208}
\author{Rastko Sknepnek}
\email{sknepnek@gmail.com}
\author{Monica Olvera de la Cruz}
\email{m-olvera@northwestern.edu}
\affiliation{Department of Materials Science and Engineering, Northwestern University,
Evanston, IL 60208}

\selectlanguage{english}%

\date{\today}
\begin{abstract}

We study closed liquid membranes that segregate into three phases due to differences in the chemical and physical properties of its components. 
The shape and in-plane membrane arrangement of the phases are coupled through phase-specific bending energies and line tensions.
We use simulated annealing Monte Carlo simulations to find low-energy structures, allowing both phase arrangement and membrane
shape to relax. The three-phase system is the simplest one in which there are multiple interface pairs, allowing
us to analyze interfacial preferences and pairwise distinct line tensions.  We observe the system's preference for interface pairs 
that maximize differences in spontaneous curvature. From a pattern selection perspective, this acts as an effective attraction 
between phases of most disparate spontaneous curvature. We show that this effective attraction is robust enough to persist 
even when the interface between these phases is the most penalized by line tension. This effect
is driven by geometry and not by any explicit component-component interaction.

\end{abstract}

\maketitle

\section{Introduction}

Multicomponent liquid membranes are pervasive in nature, and their phase behavior is important 
in many biological processes. 
Such membranes are of interest not only as constituents of complex biological systems but also as rich pattern-forming
systems in themselves. 
They provide a fertile ground for exploring the role of geometry in surface pattern formation at the nanoscale,
which is an important aspect of functional materials. 
They have been studied as basic models 
for biological systems \cite{morelli2012,chang2007artificial} and as promising candidates in the rational design
of biocompatible materials \cite{sharma1997liposomes}. Their promise is due in large part to their versatile
phase behavior. In a liquid membrane, the membrane's constituents are free to rearrange.
Although mixing is entropically favorable, this freedom has energetic consequences when there are multiple kinds of
membrane components, since different components may have different chemical and physical
properties \cite{Israelachvili1992}. When the system adapts its arrangement
to its conditions, varying shapes and component patterns can result. 

Patterns of segregated domains have been directly
observed in experimentally created multicomponent giant unilamellar
vesicles \cite{veatch2003,baumgart2003,Veatch2005ys,bacia2005sterol,Gudheti2007,delaSerna2004}.
These experiments agree with theoretical results in which patterns arise 
in response to differing bending properties of the 
membrane's segregated phases \cite{Gutlederer2009, Hu2011, gozdz2002phase,funkhouser2007coupled,harden2005budding,solis2008conditions}.
These bending properties can include bending rigidities, saddle-splay moduli, and spontaneous (preferred) 
curvatures. When these properties are phase-specific, the phase arrangement is coupled to the membrane shape.

A key aspect of surface pattern selection is the system's response to  
geometric constraint. Constraints include fixed surface area, since the energy required 
to compress a liquid membrane is orders of magnitude larger than the energy to bend it, and closure,
since tears or holes in the membrane would lead to energetically unfavorable exposure of hydrophobic tails to surrounding water. For a closed membrane
that does not exchange material with its environment, there may also be a constraint on enclosed volume. This does not apply in all cases; 
for example, lipid bilayers have been observed to adjust their volume by forming transient pores \cite{gillmor2008}. We use a constant volume 
assumption here for the sake of simplicity and note that relaxing this constraint does not qualitatively affect reported results. 
Under these constraints, a multicomponent membrane may not be able to fully indulge the preferences of all of its constituents simultaneously. 

Thus a strongly-segregated membrane may face this frustration: on the one hand, immiscibility (realized as a line tension between differing
component domains) favors macroscopic segregation and interfaces with minimal length \cite{huang1996}; on the other hand, when subject to
geometric constraints, component bending preferences may be more favorably accommodated by arrangements
with extended interfaces, including arrangements with multiple domains. In this sense it is the matter
of the length and location of interfaces that becomes the battlefront between these competing preferences.

Most studies so far have focused on two-phase systems. In two-phase
systems there is only one phase pair, hence only one type of interface. 
Introducing a third phase is a nontrivial extension.
The presence of a third phase provides three phase pairs and thus
introduces the crucial feature of interfacial preferences: 
the three-phase system has freedom not only over length and location of interfaces
but also over which phase pairs are brought into contact. 
This turns out to be an important avenue through which geometry can influence surface pattern. 
Furthermore, since the line tension between differing phase pairs
need not be the same, a three-phase system allows for an additional
type of control parameter, one which can produce qualitatively different
pattern behavior.

In this work we analyze energy-minimizing structures of closed three-phase
membranes. We observe that
system
favors interface pairs which maximize differences in 
spontaneous curvature.
A pronounced effect of this preference is
an
effective attraction between phases of most disparate spontaneous curvature, 
an indirect interaction driven by mutual response to geometric constraints. 

\begin{figure}[h]
\centering
\includegraphics[width=0.5\textwidth]{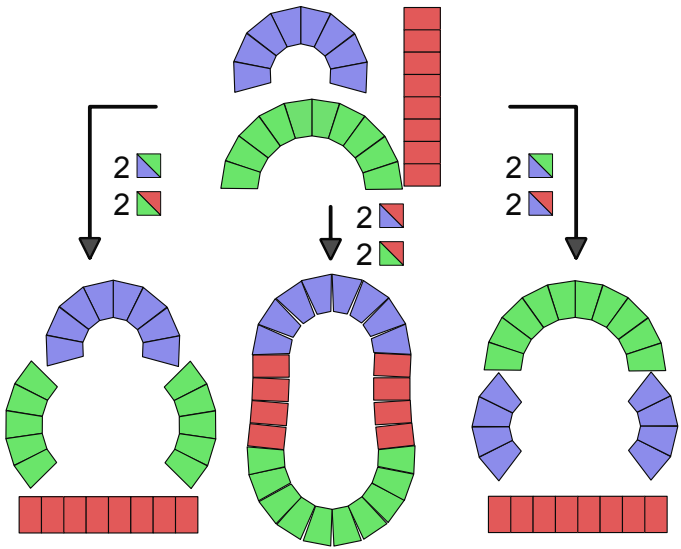}
\caption{(Color online) A cartoon representation of curvature-driven interfacial preference. Blocks of type $I$, $II$, $III$ (red, green, blue) 
represent constituents with different preferred curvatures $C_I$, $C_{II}$, $C_{III}$.
In this example $\left|C_I - C_{III}\right| > \left|C_{I}-C_{II}\right| > \left|C_{II}-C_{III}\right|$. Possible 4-interface arrangements of these constituents are shown.
The arrangement which maximizes the curvature difference at the boundaries (bottom middle) is able to
most closely accommodate its constituents' preferences when closed.}
\label{fig:cartoon}
 \end{figure}

Fig.~\ref{fig:cartoon} illustrates the basic mechanism. 
Blocks of type $I$, $II$, $III$ (red, green, blue) represent 
constituents with different preferred curvatures $C_I$, $C_{II}$, $C_{III}$.
Possible four-interface arrangements are shown.
The arrangement which maximizes the curvature difference at the boundaries is able to
most closely accomodate its constituents' preferences when closed.

We present
cases of equal line tension and cases when line tension is pairwise preferential,
showing that the geometry-induced attraction is robust enough to persist even in 
cases when it is most penalized by line tension.
This study is focused on characterizing the effective attraction. A full description of
the wide pallet of observed patterns will be published elsewhere \cite{Demers2012unpublished}.

\section{Model}

\subsection{Continuum model}

\label{sub:continuum}

We model the membrane as a two-dimensional surface. This approximation is appropriate for structures whose
lateral dimensions are much larger than the membrane thickness. The system is considered in the
strong-segregation regime, with each surface element identified with one of three phases. Each phase has a specific,
predetermined spontaneous curvature, and interfaces between
different phases are penalized with a line tension. The membrane is assumed to retain a closed spherical topology. 
We assume that the volume enclosed by the membrane is conserved, as is the area occupied by each phase.

Our simulation searches for minimum energy configurations -- that is, shape and phase arrangements which satisfy
the area and volume constraints and minimize the total energy. This energy is written as the sum of a bending
 term $\mathcal{F}_{\textrm{b}}$ and a line tension term $\mathcal{F}_{\ell}$.

The bending energy takes the form of a Helfrich functional \cite{Helfrich1973}, written as

\begin{equation}
\mathcal{F}_{\textrm{b}} = \sum\limits _{j}\int
2\kappa\left(H-C_{j}\right)^{2}\textrm{d}S_j  
\label{eq:BendingEnergy}
\end{equation}

\noindent where $j = 1,2,3$ counts phases, $H$ is the mean curvature, 
$C_{j}$ is the spontaneous curvature
of phase $j$, and $\kappa$ is the bending rigidity. 
This term penalizes surface shapes whose local curvature
deviates from its preferred value $C_j$. Since $C_j$
depends on the local composition $j$, the bending energy
couples shape and composition. Note that we do not give an index to 
bending rigidity $\kappa$ since we are assuming that the phases are equally bendable. 

Note also that the Helfrich functional 
typically includes a Gaussian curvature term weighted by saddle-splay modulus 
$\bar{\kappa}$. However, as with $\kappa$,
we are assuming that all $\bar{\kappa}$'s are equal. 
Then, since the topology of the surface
remains fixed, by the Gauss-Bonnet theorem \cite{docarmo} the Gaussian curvature term integrates to a constant
independent of the particular shape and therefore does not affect the energy.
If the $\bar{\kappa}$'s
were not equal, the term would need to be retained. Phase-specific
differences in bending rigidity $\kappa$ or saddle-splay modulus $\bar{\kappa}$ provide additional 
mechanisms for pattern formation \cite{Sknepnek2011b,Hu2011} and domain budding \cite{Yao2012}.

In monolayers, the spontaneous curvature has an intuitive physical origin
related to the properties of the head and tail groups \cite{Israelachvili1992}.
Roughly speaking, molecules with larger effective head sizes and smaller effective tail
sizes tend to pack into surfaces of negative mean curvature, much as cones would pack.
While the actual curvature properties depend on more
than steric considerations, modeling lipid types with spontaneous
curvatures has been shown to agree well with experiments \cite{gruner1989}. In lipid
bilayers, spontaneous curvature can arise from the effects of asymmetry between the inner and outer layers \cite{Svetina1989,Miao1994,dobereiner1999}.
Our model is
therefore directly applicable to monolayers, such as emulsions in which oil droplets are surrounded by
multiple surfactant types, and 
to bilayers whose phases with distinct spontaneous curvatures are conserved, such as vesicles whose inner
layer is uniform but whose outer layer is multiphase \cite{taniguchi1996}.  To extend the model to general 
multicomponent bilayers would require treatment of separate bilayer properties. 

It is worth contrasting our deformation energy with that of 
solid-like membranes.
There energy penalties associated with stretching and shear deformations can lead to 
buckling \cite{seung1988, Lidmar03} and nonlinear conformation fluctuations \cite{yoon1997}
in the homogeneous case, and in the two-component case elongated  
interconnected domains can arise \cite{vernizzi2011,Sknepnek2011b, Sknepnek2012}.
In lipid vesicles with one fluid and one solid-like phase, 
complexes of stripes and polygonal domains have been observed experimentally \cite{gordon2006}. 
These are outside the scope of our current study since none of our three phases include resistance to shear; it would be interesting to explore
interfacial preferences in these systems.

Line tension energy is given by \cite{Kohyama2003}
\begin{equation}
\mathcal{F}_{\ell} = \sum\limits _{i\neq j}\int_{\partial_{ij}} \lambda_{ij}\textrm{d}\ell
\label{eq:LineTensionEnergy}
\end{equation}

\noindent where $i,j = 1,2,3$, $\lambda_{ij}$ is the line tension between phases $i$ and $j$, 
$\textrm{d}\ell$ is a line element along $i$-$j$ interfaces, and the line integral is calculated along the boundary $\partial_{ij}$ between phases $i$ and $j$.
This term favors segregation since it penalizes interfaces between domains.

Recently a novel class of molecules called linactants has been synthesized, which behave as 2-D analogues of surfactants: they migrate laterally in a fluid layer to boundaries between immiscible phases, mediating between the immiscible constituents. Small amounts of linactant added to fluid layer have been found to reduce the line tension between immiscible phases \cite{trabelsi2008}.
This suggests that
the line tension between phases can be viewed as a tunable
design parameter.

The surface area and volume constraints can be written as $\int\textrm{d}S_j=\mathcal{A}_{j}$ ($j=1,2,3$)
and $\int\textrm{d}V=V_{0}$,
where $\mathcal{A}_{j}$ is the prescribed surface area of phase
$j$, and $V_{0}$ is the prescribed system volume. Note that there is an area constraint for each phase;
the the system's total surface area and its composition fractions are both conserved.

It is convenient to work with dimensionless parameters. Therefore, we measure energy in units of bending
rigidity $\kappa$, lengths in units of $R_0$, the radius of a reference sphere whose volume is $V_0$, curvatures in units
of $1/R_0$, and line tension in units of $\kappa/R_0$.
In these units, there remain nine independent input parameters: three spontaneous curvatures, 
three areas, and three line tensions.  While this may seem like a daunting parameter space,
we note that there is a great deal of symmetry and physical analogy, and we believe our choice of
parameter pairs for simulation is representative and can provide useful insight into the phenomena at hand.

\subsection{Discrete model}

\label{sub:discrete}

Our numerical method is based on a triangulation of the surface.
Each vertex is identified with a phase type and surface area elements are represented by polygonal patches
centered at vertices. The discrete analogue of the bending energy, Eq.~\eqref{eq:BendingEnergy},
is computed as a sum over vertices, with the discrete mean curvature
near each vertex $v$ computed as \cite{meyer},

\begin{equation}
H_{\textrm{discrete}}(v) = \frac{\frac{1}{4}\sum\limits_{i}\left|e_i\right|\psi_{i} }{\sum\limits_{j}\frac{1}{3}A_j},
\label{eq:Discrete_mean_curvature}
\end{equation}

\noindent where $i$ indexes edges for which $v$ is an end point, $\left|e_i\right|$ is the length of the $i^{\textrm{th}}$ edge,
$\psi_{i}$ is the dihedral angle between two triangles sharing the $i^{\textrm{th}}$ edge, $j$ indexes triangles containing $v$, and
$A_j$ is the area of the $j^{\textrm{th}}$ triangle.

To compute discrete line tension, Eq.~\eqref{eq:LineTensionEnergy},
we sum over all adjacent vertex pairs of different phase types \cite{Kohyama2003}:
$\sum\limits _{\left\langle i,j\right\rangle }\lambda_{ij}\left(\left|\textbf{c}_{ij}^{(1)}-\textbf{m}_{ij}\right|+\left|\textbf{c}_{ij}^{(2)}-\textbf{m}_{ij}\right|\right)$, 
where $\left\langle i,j\right\rangle $ denotes pairs of
adjacent vertices $i$ and $j$ that have different phase types, $\textbf{c}_{ij}^{(1)}$
and $\textbf{c}_{ij}^{(2)}$ are the centers of the two triangles
that have $i$ and $j$ as vertices, and $\textbf{m}_{ij}$ is the
midpoint of their shared edge. Note that this differs from the approach used by Hu, \emph{et al.}~\cite{Hu2011}, as
we take into account local triangle distortions when computing interface length.

The surface area constraint is enforced by penalizing differences between $A(i)$, the surface area associated
with a vertex $i$, and the prescribed per-vertex value,
$\mathcal{F}_{\textrm{a}}=\sum\limits _{i}\frac{\sigma}{2}\left(A(i)-A_{0}(i)\right)^{2}$.
The volume constraint is also enforced by a harmonic potential penalty.

Throughout all simulations, we set $\sigma=2\times10^{5}$ and $\nu=10^{4}$.
These values were tuned to ensure that, by the end of a run, the total
volume and surface areas differed from their target values by less
than $1\%$. Using $\phi_{i}$ to denote the composition fraction of phase $i$,
the prescribed surface area of phase $i$ is $1.05\times4\pi\phi_{i}$.
This is $5\%$ larger than the area
of a sphere of identical volume, excess being necessary to
avoid the trivial case of an undeformable sphere.

To find low-energy configurations, simulated annealing Monte Carlo simulations were performed
following a linear cooling protocol. Three types of Monte Carlo moves were used: (i) a surface move,
which attempts to perturb the position of a vertex by $\left|\Delta\vec r\right|=2.5\times10^{-3}$, (ii) a phase-swap
move, which attempts to exchange the phase type for two randomly
selected vertices, and (iii) an edge flip, which cuts an edge shared by two triangles and reattaches it
so that it spans the opposite previously-unattached vertices \cite{Kazakov1985,Billoire1986}. All moves were accepted according to the Metropolis
rules. $4\times10^{6}$ sweeps, with a sweep defined as an attempted move of each vertex,
were performed for each parameter set. Phase swap moves were performed every five and edge flip moves every ten sweeps. 
Initial configurations were random triangulations of a sphere, constructed from regular, Caspar-Klug triangulations \cite{Caspar62} such that vertex moves
were constraint to a sphere and edge flips were performed at a very high temperature to ensure high acceptance rates. Any moves or edge flips that would result in 
unphysical edge crossings were rejected. The resulting configuration had a large number of vertices with coordination different than six.
Finally, vertex types were randomly permuted. Each vertex was assumed to have a hard core of diameter
$l_{min}=0.093$, and each edge was endowed with a tethering potential with maximum length $l_{max}=0.157$ such that $l_{max}/l_{min}=1.688$.
These values were chosen to be tight enough to prevent membrane self-intersection 
but slack enough to allow edge-flips \cite{Gompper2000,Kohyama2003}.

Since the system's energy landscape is complicated, the computed configurations are not guaranteed 
to represent global minima. However, because independent runs starting from different random 
initial configurations reproduce the same qualitative features, 
we can regard the results as \emph{typical}. 

Patterns are classified using graphs: 
each colored domain is identified with a colored node, and two nodes are linked if they share an interface. 
Then two configurations are considered to represent the same pattern if their 
graphs are isomorphic, respecting color.

\section{Results and Discussion}

\begin{figure}[htb]
\centering
\includegraphics[width=0.5\textwidth]{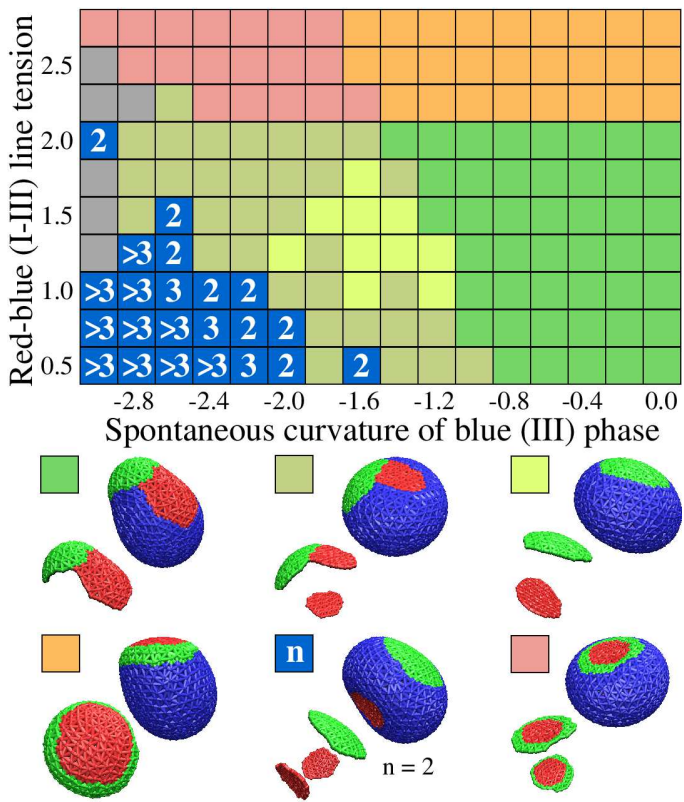}
\caption{(Color online) Phase diagram for composition ratio $I:II:III = \frac{1}{10}:\frac{1}{10}:\frac{8}{10}$,
varying spontaneous curvature $C_{III}$ and line tension $\lambda_{I,III}$. All other parameters are fixed, 
with $C_I = 0.0$, $C_{II} = -1.0$, and $\lambda_{I,II} = \lambda_{II,III} = 1.0$. Phases $I,II,III$ are
shown in red (medium), green (lightest), and blue (darkest), respectively. Each example vesicle is shown twice:
on the upper right, all phases are visible; on the lower left, one phase has been removed so that the pattern can be more clearly seen. 
Gray squares denote assorted structures with multiple green and red domains. All snapshots were generated with the Visual 
Molecular Dynamics (VMD) package \cite{Humphrey96} and rendered with the Tachyon ray-tracer \cite{Stone98}.}
\label{fig:NoMatrices_VaryLT_101080}
\end{figure}

A sample of the varied patterns is provided in Fig.~\ref{fig:NoMatrices_VaryLT_101080},
where we present a phase diagram for a 
$\frac{1}{10}:\frac{1}{10}:\frac{8}{10}$ mixture of components
$I:II:III$ (red:green:blue). 
In this diagram, we vary $C_{III}$ (spontaneous curvature of the blue phase)
and $\lambda_{I,III}$ (line tension between red and blue phases).
We fix spontaneous curvatures
$C_I=0$, $C_{II}=-1.0$ and line tensions $\lambda_{I,II}=\lambda_{II,III}=1.0$.
Results for each parameter set are classified according to pattern. We
define pattern in terms of the number and arrangement of colored domains,
irrespective of shape.  

For present purposes, we note two features. The first is that a preferential line tension opens the possibility
of qualitatively different pattern behavior and thus can be an important control parameter
in the system's pattern selection. Furthermore, we observe a tranisition near $C_{III}=-1.0$,
which is the dividing line where the spontaneous curvature of phase $III$ switches from middle to largest in magnitude. 

\begin{figure}[h]
\centering
\includegraphics[width=0.5\textwidth]{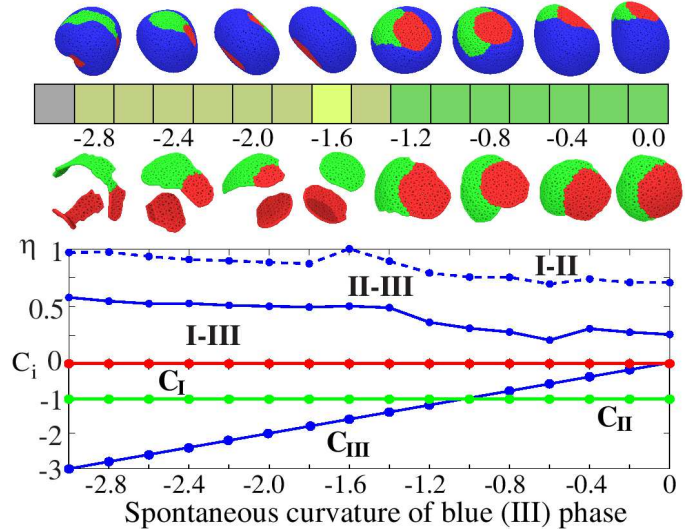}
\caption{(Color online) Analysis of a row of the diagram shown in Fig.~\ref{fig:NoMatrices_VaryLT_101080} for $\lambda=1.75$. 
Composition fraction is set to $\frac{1}{10}:\frac{1}{10}:\frac{8}{10}$ and $C_{III}$ is varied; 
$C_{I} = 0.0$, $C_{II}=-1.0$, $\lambda_{I,II} =\lambda_{II,III}=1.0$ are kept fixed.
\emph{Upper:} Configurations with colored squares indicating pattern grouping. Phases $I,II,III$ are
shown in red (medium), green (lightest), and blue (darkest), respectively.
\emph{Middle:} Length of each interface type to total interface length
ratio $\eta$ as a function of $C_{III}$. Interface fraction is represented by height between curves. 
\emph{Bottom:} Spontaneous curvatures, with red, green, and blue lines corresponding to spontaneous curvatures of phases of
type $I$, $II$, and $III$, respectively. }
\label{fig:curvature_comparison_101080}
 \end{figure}

In Fig.~\ref{fig:curvature_comparison_101080} we analyze in detail the $\lambda_{I,III} = 1.75$ row from the diagram in
 Fig.~\ref{fig:NoMatrices_VaryLT_101080}. 
 Note that $I-III$ interfaces predominate in the region where
 $C_{III}<C_{II}$, even though such interfaces are nearly twice as costly as $I-II$ and $II-III$ interfaces. 

\begin{figure}[h!]
\centering
\includegraphics[width=0.5\textwidth]{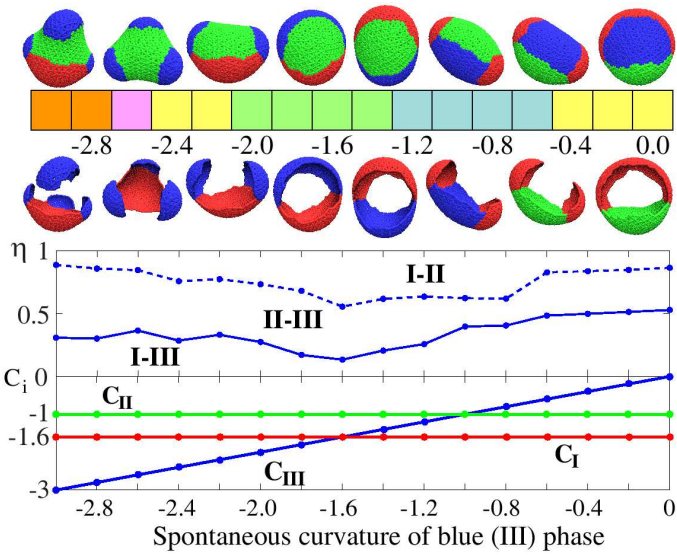}
\caption{\emph{Upper:} Configurations with colored squares indicating pattern grouping. Phases $I,II,III$ are
shown in red (medium), green (lightest), and blue (darkest), respectively.
\emph{Middle:} Length of each interface type to total interface length
ratio $\eta$ as a function of $C_{I}$. Interface
fraction is represented by height between curves. \emph{Lower:} Spontaneous curvatures,
with red, green, and blue lines corresponding to spontaneous curvature of phases $I$, $II$, and $III$, respectively. The composition
fraction is set to $\frac{1}{3}:\frac{1}{3}:\frac{1}{3}$ and $C_{I}=-1.6$, $C_{II}=-1.0$, $\lambda_{I,II}=\lambda_{II,III}=\lambda_{I,III}=1.0$ 
are kept fixed. 
}
\label{fig:curvature_comparison_33333}
\end{figure}

Fig.~\ref{fig:curvature_comparison_33333} examines a $\frac{1}{3} : \frac{1}{3} : \frac{1}{3}$ composition ratio where all line tensions are set to $1.0$, 
with $C_{I} = -1.6$ and $C_{II}=-1.0$. In this row we observe that the spontaneous curvature of phase $III$ (blue) transitions
from smallest in magnitude to middle to largest. Green-blue ($II-III$) interfaces are
most favored on the left, where blue is largest in magnitude and green is smallest. In the middle, green ($II$) is smallest in magnitude and red ($I$) and blue ($III$) are close; here red-green ($I-II$) and blue-green ($II-III$) interfaces are about equally favorable, and red-blue ($I-III$) interfaces are disfavored.  On the right, blue is largest in magnitude and red is smallest, and red-blue interface is most favored.  Red and green are closest in magnitude, and red-green interfaces are disfavored.  The interfaces between phases of most disparate spontaneous curvature are most favored.

\begin{figure}[h!]
\centering
\includegraphics[width=0.5\textwidth]{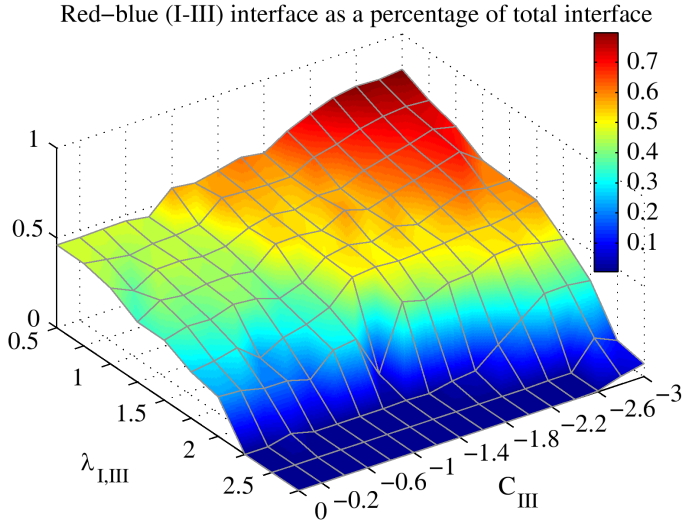}
\caption{(Color online) Length of $I-III$ (red-blue) interface as a fraction of total interface, for composition ratio $I:II:III = \frac{1}{10}:\frac{1}{10}:\frac{8}{10}$, varying spontaneous curvature $C_{III}$ and line tension $\lambda_{I,III}$.
All other parameters are fixed, with $C_I = 0.0$, $C_{II} = -1.0$, and $\lambda_{I,II} = \lambda_{II,III} = 1.0$. }
\label{fig:RBInterface_VaryLT}
 \end{figure}

\begin{figure}[thb]
\centering
\includegraphics[width=0.5\textwidth]{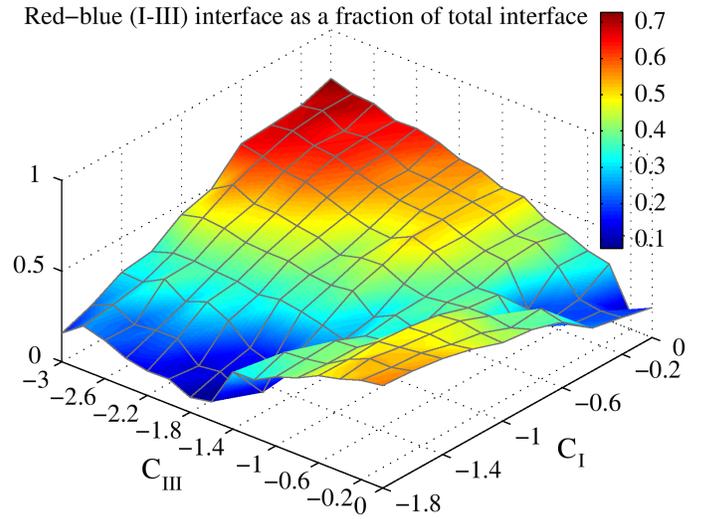}
\caption{(Color online) Length of $I-III$ (red-blue) interface as a fraction of total interface, for composition ratio
$I:II:III = \frac{1}{3}:\frac{1}{3}:\frac{1}{3}$, varying spontaneous curvatures $C_I$ and $C_{III}$. 
All other parameters are fixed. $C_{II} = -1.0$, and $\lambda_{I,II} = \lambda_{I,III} = \lambda_{II,III} = 1.0$. }
\label{fig:RBInterface_FixLT_333333}
\end{figure}

In Figs.~\ref{fig:RBInterface_VaryLT} and \ref{fig:RBInterface_FixLT_333333}, we show that these interfacial preferences hold over wide slices of parameter space.
Fig.~\ref{fig:RBInterface_VaryLT} shows $I-III$ interface length as a fraction of total interface for
the configurations obtained from the parameter sets used in Fig.~\ref{fig:NoMatrices_VaryLT_101080}. In the regime
where $C_{III}<C_{II}<C_{I}$, $I-III$ interfaces predominate until they become twice as energetically expensive as $I-II$ or $II-III$. 

Fig.~\ref{fig:RBInterface_FixLT_333333} plots $I-III$ interface fraction
in the system with $\frac{1}{3}:\frac{1}{3}:\frac{1}{3}$ composition ratio, 
varying $C_{I}$ and $C_{III}$. $\lambda_{I,II}=\lambda_{I,III}=\lambda_{II,III}=1.0$ and $C_{II}=-1.0$ remain fixed.
As expected, symmetry about the line $C_{I}=C_{III}$ is evident. We observe the fraction of the $I-III$ interface 
predominating in regions where $C_{I}>C_{II}>C_{III}$ or $C_{I}<C_{II}<C_{III}$. Conversely, $I-III$ interfaces are strongly disfavored in regions where $\left|C_{I}-C_{III}\right| < \left|C_{I}-C_{II}\right|$ and $\left|C_{I}-C_{III}\right| < \left|C_{II}-C_{III}\right|$.

We point out that for large values of line tension or spontaneous curvatures, well beyond the range used in this study, 
one observes interesting budding effects. Budding plays an important role in biological systems \cite{Pornillos2002,Bonifacino2004,Welsch2009} and has been investigated 
in a number of studies \cite{Julicher1993,Miao1994,Julicher1996,Kumar2001,Kohyama2003,Laradji2005,harden2005budding}. Effects of interfacial preference on budding in a three-phase
liquid membrane will be addressed elsewhere.

In conclusion, we have seen that mutual response to geometry
acts as an effective attraction between phases of most disparate
spontaneous curvature in a three-phase liquid membrane. This effect arises indirectly through the coupling of deformation
and compositional arrangement, rather than through a direct component-component interaction.
Nonetheless, it is robust enough to compete with a countervailing line tension. In some cases it results in
predominance of interfaces between phases least miscible by pure line tension considerations. Therefore, this system 
provides an example where geometric constraints, rather than direct interactions, can dominate its conformation. 
Our findings suggest that an intricate interplay between the geometry and composition can lead to a rich phase behavior of 
complex fluid membranes. We hope that our results will stimulate further experimental and theoretical work on these rich systems.

We would like to thank S.~Patala, F.~Solis, and C.K.~Thomas for useful discussions.
Numerical simulations were in part performed using the
Northwestern University High Performance Computing Cluster Quest.
MFD and MO thank the financial support of the Air Force Office
of Scientific Research (AFOSR) under Award No. FA9550-10-1-0167.
RS and MO thank the financial support of the US Department of
Energy Award DEFG02-08ER46539 and the Office of the Director of
Defense Research and Engineering (DDR\&E).

\bibliographystyle{apsrev}

\end{document}